# Can classical physics agree with quantum physics on quantum phenomena?

Michele Marrocco

*ENEA*

*(Italian National Agency for New Technologies, Energies and Sustainable Economic Development)*

*via Anguillarese 301, Rome, 00123 Italy*

*michele.marrocco@enea.it*

Classical physics fails where quantum physics prevails. This common understanding applies to quantum phenomena that are acknowledged to be beyond the reach of classical physics. Here, we make an attempt at weakening this solid belief that classical physics is unfit to explain the quantum world. The trial run is the quantization of the free radiation field that will be addressed by following a strategy that is free from operators or quantum-mechanical concepts.

The intimate laws of nature fall under the realm of quantum physics [1-3]. Classical physics seems to be out of that territory controlled by quantum laws, despite the fact that most of our knowledge (and most of our world) is classical. This divide dates back to those years when new methods were conceived to capture the meaning of what could not be explained in terms of classical concepts. However, the question of whether the divide between quantum and classical physics is smaller than it seems is stimulating and challenging.

The question motivates the current work. In particular, we consider one of the most known problems in quantum mechanics, the quantization of the free radiation field [1-3]. The quantization is usually realized according to the correspondence principle of quantum mechanics: classical physics should be recovered by the quantum laws under the macroscopic limit. It is for this reason that we can set up the correspondence that connects a series of harmonic oscillators to the free radiation field. The procedure is very popular and can be found in any textbook on the subject. The result is that the energy of the radiation field is $(n+1/2)\hbar\omega$ where $n$ is an integer that takes the meaning of photon number, $\hbar$ is the reduced Planck constant and $\omega$ is the angular frequency of the field.

In this work, we want to reproduce this fundamental law by means of a procedure that is free from traditional quantum-mechanical concepts. The key idea is related to the classical multipole expansion of the electromagnetic field [4] and, surprisingly, although we do not rely on the help provided by the quantum harmonic oscillator, we are able to find the quantum-mechanical result provided that we keep ourselves from establishing an immediate connection of the running index $n$ to the photon number. To this end, the term of energy quanta will be preferred in reference to the procedure illustrated in the current work, whereas the term photons remains for the quantum-mechanical procedure of quantization.

The alternative quantization procedure starts where the quantum correspondence breaks in. That is at the point where the vector potential is used to establish the correspondence of the cycle-averaged classical energy with the Hamiltonian of the multitude of quantum oscillators. In general, the classical-averaged energy is written as

$$<H> = \frac{1}{2}\varepsilon_0 \int d\mathbf{r} <|\mathbf{E}(\mathbf{r},t)|^2> + \frac{1}{2\mu_0}\int d\mathbf{r} <|\mathbf{B}(\mathbf{r},t)|^2> \quad (1)$$

where $\mathbf{E}(\mathbf{r},t)$ and $\mathbf{B}(\mathbf{r},t)$ are the electric and magnetic fields whereas the electric permittivity and magnetic permeability of free space are related to the speed of light $c = 1/\sqrt{\varepsilon_0\mu_0}$. However, in place of Eq. (1), we are going to work on the simpler expression

$$<H> = \varepsilon_0 \int d\mathbf{r} <|\mathbf{E}(\mathbf{r},t)|^2> \quad (2)$$

that involves the known balance between the electric term and its magnetic counterpart.

Next, we introduce the mode summation

$$\mathbf{E}(\mathbf{r},t) = \sum_{\mathbf{k}} \mathbf{E}_{\mathbf{k}}(\mathbf{r},t) = \sum_{\mathbf{k},s} E_{\mathbf{k},s}(\mathbf{r},t)\mathbf{e}_{\mathbf{k},s} \quad \text{with} \quad \mathbf{E}_{\mathbf{k}}(\mathbf{r},t) = \sum_{s} E_{\mathbf{k},s}(\mathbf{r},t)\mathbf{e}_{\mathbf{k},s} \quad (3)$$

plus the fact that we can decompose Eq. (11) into mode contributions

$$<H> = \varepsilon_0 \sum_{\mathbf{k},s} \int d\mathbf{r} <|E_{\mathbf{k},s}(\mathbf{r},t)|^2> \quad (4)$$

where the components $E_{\mathbf{k},s}(\mathbf{r},t)$ of the electric field satisfy their own wave equation

$$\nabla^2 E_{\mathbf{k},s}(\mathbf{r},t) = \frac{1}{c^2}\frac{\partial^2 E_{\mathbf{k},s}(\mathbf{r},t)}{\partial t^2} \quad (5)$$

The scheme behind Eqs. (2)-(5) is manifest. Unlike the quantum approach where everything revolves around the vector potential, we prefer to handle the true observable, that is, the electric field. In addition, we avoid the plane-wave expansion typical of the quantum-mechanical procedure. This choice reduces the whole problem to a specific form of wave propagation. Our choice is different and explained below. Nonetheless, we keep the fact that the double time derivative transforms the right-hand side of Eq. (5) into $-k^2 E_{\mathbf{k},s}(\mathbf{r},t)$, having assumed that $\omega_k = ck$. With these preliminaries, we are prepared for the procedure that does not require quantum-mechanical operators.

The procedure is not totally new. It is partially taken from the classical multipole expansion of electromagnetic fields [pag. 429 of Ref. 4], according to which the spatial dependences are encoded in the well-known Helmholtz equation

$$\nabla^2 E_{\mathbf{k},s}(\mathbf{r},t) + k^2 E_{\mathbf{k},s}(\mathbf{r},t) = 0 \tag{6}$$

which is familiar to many in mathematics and physics.

The general solution of Eq. (5) is achieved by introducing polar coordinates $r$, $\varphi$ and $\vartheta$ and contains special functions. The radial component is given by spherical Bessel and Neumann functions [pag. 425 of of Ref. 4]. The angular dependence is instead determined by the spherical harmonics that form an orthonormal basis in the Hilbert space of square-integrable functions [pag. 108 of Ref. 4]. These special functions appear in classical electrodynamics [4], classical optics [5], acoustics [6], geophysics [7] and beyond these examples, they are central to the quantum-mechanical determination of orbital angular momenta (see Ref. 1, pp. 519-523). Soon we will discover that the spherical harmonics are central to this demonstration too.

The demonstration is given for one solution of Eq. (6). The solution is chosen regular at the origin (we are in the free space, i.e, without singularities, and the field has to be regular!). This means that the solution can be written in dependence on the generic order $n$ of the spherical Bessel function $j_n(kr)$

$$E_{\mathbf{k},s}(r,\vartheta,\varphi,t) = E_0(t)\, j_n(kr) \sum_{m=-n}^{n} Y_n^m(\vartheta,\varphi)\ . \tag{7}$$

with the $k$ dependence incorporated in the radial component only. The amplitude $E_0(t)$ is time dependent. Since no charges are within the region where the integral of Eq. (4) is calculated, the Maxwell's condition $\nabla \cdot \mathbf{E_k} = 0$ implies that $E_0(t)$ does not depend on the order $n$ of the spherical Bessel function. The amplitude $E_0(t)$ is also assumed independent from the polarization because of the free choice we have on the transverse polarization vectors $\mathbf{e}_{\mathbf{k},s}$ of $\mathbf{E_k}$ in Eq. (3). However, regardless of these secondary details, our argument touches the first delicate point. Combining Eq. (7) with Eq. (4) The integration over the solid angle becomes

$$\sum_{m,m'=-n}^{n} \int\int d\vartheta d\varphi \sin\vartheta \, Y_n^{m'*}(\vartheta,\varphi) \, Y_n^m(\vartheta,\varphi) = 2n+1 \qquad (8)$$

that brings out something very interesting on the right-hand side. The orthonormality condition for $Y_n^m(\vartheta,\varphi)$ (see Eq. 3.55, page 108, in Ref. 4) produces a degeneracy of $2n+1$ terms in the calculation of the energy. This result goes with the cycle average of the square of the time dependent amplitude

$$<|E_0(t)|^2> = \frac{|E_0|^2}{2} \qquad (9)$$

that generates a time-independent factor $|E_0|^2$ and a factor of $1/2$ [coming from the cycle average of $sin^2(\omega t)$ or $cos^2(\omega t)$ terms]. In conclusion, our calculation of the mode energy is

$$\varepsilon_0 \int d\mathbf{r} <|E_{\mathbf{k},s}(\mathbf{r},t)|^2> = \varepsilon_0 |E_0|^2 R_n \left(n+\frac{1}{2}\right) \qquad (10)$$

with the radial integral $R_n$ written as

$$R_n = \int_0^R dr \, r^2 \, j_n^2(kr). \qquad (11)$$

where $R$ is the radius of the quantization volume $V$ that is also present in the classical-quantum correspondence that creates the well-known quantum-mechanical result for the energy.

It is undeniable that Eq. (10) has something that reminds us of the quantum-mechanical result. We disregard he radial integral $R_n$ at the moment. We anticipate that its dependence on $n$ is very weak and disappears when we consider the limit of $kR >> 1$. In other words, we have found the $n+1/2$ rule by applying a simple procedure that is well known in electrodynamics [4]. The difference with the quantum-mechanical procedure is on the meaning of the number $n$. In the quantum-mechanical approach, $n$ is the eigenvalue of the photon number operator $\hat{a}^+_{\mathbf{k},s}\hat{a}_{\mathbf{k},s}$. In the current attempt, $n$ is the number of spherical harmonics of opposite secondary index $m$. We do not want to make a direct connection to the photon number, nevertheless, this classical procedure finds a quantization rule that looks like the quantum-mechanical rule.

There is something more besides the similarity of the $n+1/2$ rule. It concerns the vacuum field. Indeed, the most prominent result of Eq. (10) is what could be called the classical vacuum field connected to the zero-point energy for $n=0$. This condition leaves room for the fundamental ($m=0$) spherical harmonic only (monopole term). Its complete spherical symmetry supports the view that the vacuum field is spatially isotropic and, by looking at Eq. (8), its contribution to the energy counts for one unit instead of the elusive half-photon of the quantum-mechanical approach. The value of $1/2$ in Eq. (10) is, indeed, accidental because its appearance is caused by the cycle average of Eq. (9) and has nothing to do with the mysterious energy splitting suggested by the quantum-mechanical understanding of the vacuum energy.

Furthermore, Eq. (10) identifies a peculiar value of reference for the elementary electric field amplitude $E_0$ associated with each energy quantum. In this manner, we can picture the energy of the electromagnetic field in relation to a unique value of the electric field. By contrast, higher energy levels are not given by a larger value of $E_0$. Higher energy levels are realized by the excitation of spherical harmonics of higher order than the fundamental. This is in marked contrast with what is believed to characterize the classical view of the radiation field and we have found that a pure classical argument agrees with quantum mechanics on the idea of what energy is.

Another consequence of Eq. (10) is relative to the concept of state. For instance, the vacuum state of $n = 0$ and $m = 0$ corresponds to the monopole term and this implies that any point in the space is a source of the vacuum. Next, the one-particle state of $n = 1$ corresponds to a state of the first three harmonics with $m = 0$ and $m = \pm 1$ (dipole) as if the energy quanta had an internal structure consisting in their own spin. This is not far from the quantum-mechanical picture. The state with $n = 2$ corresponds to a state of the first five harmonics with $m = 0$ and the possible combination of $m = \pm 1, \pm 2$ (quadrupole) and so on for $n \geq 3$.

The final image we have from what has been accomplished so far is rather promising and we go on with the calculation.

The procedure leading to Eq. (10) is very simple and hinges on the solution of the Helmholtz equation for the electric field in place of plane-wave expansions of the vector potential. However, we did not take into proper account the problematic role of the radial integral $R_n$. We do it now.

The radial part of Eq. (7) is made of multipole contributions that multiply spherical waves. It means that, when $R_n$ is calculated, we obtain a term that depends linearly on the radius $R$ plus oscillating terms. The result can be written in closed form, but we are especially concerned about the limit of $kR \gg 1$ that corresponds to the cases of practical interest. Fortunately, the great value of $kR$ suppresses the oscillating terms and we find that $R_n$ is independent from $n$. In this case, being $R_n = R/(2k^2)$, Eq. (10) becomes

$$\int d\mathbf{r} <|E_{\mathbf{k},s}(\mathbf{r},t)|^2> = \left(n_{\mathbf{k},s} + \frac{1}{2}\right)|E_0|^2 \frac{R}{2k^2}. \tag{12}$$

where we have made explicit the obvious connection of the number $n_{\mathbf{k},s}$ of energy quanta to the mode $(\mathbf{k},s)$ of the field of which we calculate the integral on the left-hand side. One last step is still missing to obtain the energy of the electromagnetic field for a specific angular frequency (monochromatic limit). We need to count the modes associated with the chosen condition of one single frequency. The modes are counted according to the Rayleigh-Jeans technique and are $N_{mod} = Vk^3/(3\pi^2)$ for the reference volume $V$ of quantization [8]. In this manner, the summation of Eq. (4) produces $N_{mod}$ that multiplies Eq. (12) and the final result for each state of polarization is

$$<H_s> = \beta\left(n_{\omega,s} + \frac{1}{2}\right)\omega \qquad \text{with} \quad \beta = \frac{\varepsilon_0}{6\pi^2 c} RV|E_0|^2 \tag{13}$$

It is now easy to observe that Eq. (13) is nothing but the quantum-mechanical expression except for the parameters collected in $\beta$ and the meaning of $n_{\omega,s}$. We might also dream of reaching the conclusion that $\beta$ equals the reduced Planck constant $\hbar$. However, we avoid this.

**REFERENCES**

[1] A. Messiah, *Quantum Mechanics* (Dover Publications, New York, 1999).

[2] M. O. Scully and M. S. Zubairy, Quantum Optics (Cambridge University Press, Cambridge, 1999).
[1] A. Messiah, *Quantum Mechanics* (Dover Publications, New York, 1999).

[2] M. O. Scully and M. S. Zubairy, Quantum Optics (Cambridge University Press, Cambridge, 1999).